\documentclass[aps,prl,twocolumn,groupedaddress,showpacs,amsmath,amssymb,floatfix]{revtex4}
\usepackage{graphicx,bm}

\begin{document}

% number in the upper righthand corner of the title page in preprint
% mode.
%\preprint{}

\title{ Dynamical density functional theory with hydrodynamic interactions and
colloids in unstable traps}


\author{M. Rex}
\email{rexm@thphy.uni-duesseldorf.de}
%\thanks{}
\affiliation{Institut f\"ur Theoretische Physik II: Weiche Materie,
Heinrich-Heine-Universit\"at D\"{u}sseldorf, 
Universit{\"a}tsstra{\ss}e 1, D-40225 D\"{u}sseldorf,
Germany}
\author{H. L{\"o}wen}
\affiliation{Institut f\"ur Theoretische Physik II: Weiche Materie,
Heinrich-Heine-Universit\"at D\"{u}sseldorf, 
Universit{\"a}tsstra{\ss}e 1, D-40225 D\"{u}sseldorf,
Germany}

\date{\today}

\begin{abstract}
A density functional theory for colloidal dynamics is presented which
includes hydrodynamic interactions between the colloidal particles.
The theory is applied to the dynamics of colloidal particles in an
optical trap which switches periodically in time from a stable to
unstable confining potential. In the absence of hydrodynamic interactions,
the resulting density breathing mode, exhibits huge
oscillations in the trap center which are almost completely
 damped by hydrodynamic interactions.
The predicted dynamical density fields are in good agreement with
Brownian dynamics computer simulations.
\end{abstract}


\maketitle

\section{Introduction}

The dynamics of mesoscopic colloidal particles dispersed in a
molecular solvent is of fundamental importance for an understanding of
soft matter transport and flow properties.
The control over the collective colloidal dynamics leads to the
construction of "smart" materials steered by external fields like
electro- or magnetorheological fluids \cite{ma_ap_2003} and is essential
for applications such as gelation and aggregation in paints and
cosmetics \cite{sciortino_ap_2005}.
Apart from the stochastic Brownian motion of the colloidal particles
due to their kicks with the solvent molecules, hydrodynamic
interactions between colloidal particles arising from the induced
solvent flow field are getting relevant for concentrated suspensions.
It has been shown by experiments \cite{banchio_prl_2006,lutz_epl_2006}, computer
simulations \cite{padding_prl_2004,kuusela_prl_2003} and theory
\cite{naegele_jcp_1998_a,naegele_jcp_1998_b,naegele_jcp_1999,cichocki_jcp_2002} 
that hydrodynamic interactions can lead to
qualitatively different behavior in the bulk transport properties and
in colloidal sedimentation as compared to simple Brownian motion valid
at very low volume fractions.

A full microscopic theory which starts from the colloidal interactions
and their hydrodynamic mobility tensors and predicts the dynamical
properties is in principle possible by starting from the Smoluchowski
picture \cite{dhont_book}.
In practice, however, such a predictive theory is hampered by the
many-body nature of the problem and the long range of the Oseen
mobility tensor which is the leading contribution for a colloidal
pair.
Explicit approaches have been worked out in the bulk for short-time
and long-time diffusion coefficients
\cite{naegele_jcp_1998_a,naegele_jcp_1998_b,naegele_jcp_1999}, and for the
viscosity \cite{naegele_jpcm_2003}.
There are also first investigations for colloids near walls and on
interfaces \cite{cichocki_jcp_2002,pesche_epl_2000}, and for the
nonequilibrium structure of colloids \cite{lionberger_jr_1997} but a
general theory for an arbitrary and time-dependent inhomogeneous
external potential is missing.

The goal of this letter is twofold: first we construct a dynamical
density functional theory \cite{chan_prl_2005} which incorporates
hydrodynamic interactions.
The theory is explicitly worked out for hydrodynamic interactions on
the Rotne-Prager level and generalizes earlier formulations
\cite{archer_jcp_2004,tarazona_jcp_1999,tarazona_jpcm_2000} where
hydrodynamic interactions were neglected.
The theory makes predictions for an arbitrary time-dependent external
potential, i.e., for a general inhomogeneous nonequilibrium situation.
Second, we apply the theory to the dynamics of colloidal particles
confined in an unstable optical trap which switches periodically in
time from a stable to unstable confining potential.
This situation can in principle be realized, e.g., by combining two
laser tweezers or by scanning around a single laser tweezer quickly
\cite{henderson_prl_2002,martin_prl_2006,lutz_epl_2006}.
The response to this oscillating trap is a time-dependent radial-symmetric 
one-particle density profile which we call - in analogy to
trapped Bose gases \cite{griffin_prl_1997} - a driven {\it breathing mode}.
The periodic breathing mode is interesting in
itself since it may serve as a hydrodynamic transmitter
\cite{guidarelli_us_1998}. 

As a result
the properties of the breathing mode strongly depend on hydrodynamic
interactions. For instance, significant oscillations in the density profile
in the trap center which built up if no hydrodynamic interactions are present
are completely damped by hydrodynamic interactions.
The predictions of the dynamical density functional theory are in very
good agreement with Brownian dynamics nonequilibrium computer
simulations which take hydrodynamic interactions into account on the
same level as the theory does.
The theory can in principle be applied to any inhomogeneous situation
like laser-induced freezing \cite{chowdhury_prl_1985}.
It may even be a possible route to incorporate hydrodynamic
interactions into dynamical approaches like mode-coupling theory since
the latter can be brought into relation with dynamical density functional theory
\cite{archer_jpcm_2006}.

The starting point of the derivation of the dynamical density
functional theory including hydrodynamic interactions on the two-body
level is the Smoluchowski equation, i.e., the equation for the
time-evolution of the full probability density distribution
$P(\mathbf{r}^{N},t)$ for $N$ interacting spherical Brownian particles
at positions
$\mathbf{r}^{N}=\mathbf{r}_{1},\mathbf{r}_{2},...,\mathbf{r}_{N}$ and time $t$ (see
e.g.\ \cite{dhont_book}):
\begin{equation}
  \label{eq:hismoluchowski}
    \frac{\ensuremath{\partial{P(\mathbf{r}^N,t)}}}{\ensuremath{\partial{t}}}
    = \sum_{i,j}^{N}%\sum_{j=1}^{N}
    \mathbf{\nabla}_{i}\cdot\mathbf{D}_{ij}(\mathbf{r}^{N})
    \cdot\left[ 
    \mathbf{\nabla}_{j} +
    \frac{\mathbf{\nabla}_{j}U(\mathbf{r}^{N},t)}{k_{\mathrm{B}}T}
    \right ]P(\mathbf{r}^{N},t),
\end{equation}
where $k_{\mathrm{B}}T$ is the thermal energy.
We assume pairwise additivity for the total potential energy
of the system, such that $U(\mathbf{r}^{N},t) = \sum_{k=1}^{N}
V_{\mathrm{ext}}(\mathbf{r}_{k},t) + \frac{1}{2}\sum_{k=1}^{N}\sum_{l
\neq k}^{N} v_{2}(\mathbf{r}_{k},\mathbf{r}_{l})$, where
$V_{\mathrm{ext}}(\mathbf{r},t)$ is the one-body time-dependent
external potential acting on each particle and
$v_{2}(\mathbf{r},\mathbf{r}^{\prime})$ is the pair interaction
potential.
Hydrodynamic interactions are included through the
configuration-dependent diffusion tensor which we approximate on a two
particle level: $\mathbf{D}_{ij}(\mathbf{r}^{N})\approx
D_{0}\mathbf{1}\delta_{ij} + D_{0}\left [ \delta_{ij}\sum_{l\neq
i}^{N} \bm{\omega}_{11}(\mathbf{r}_{i}-\mathbf{r}_{l}) +
(1-\delta_{ij}) \bm{\omega}_{12}(\mathbf{r}_{i}-\mathbf{r}_{j})\right
]$. Here,
$D_{0}$ denotes the diffusion constant of a single isolated particle
and $\delta_{ij}$ is Kronecker's delta.
For a one-component suspension of spheres, series expansions of the two
tensors $\bm{\omega}_{11}$ and $\bm{\omega}_{12}$ are known, in principle,
to arbitrary order \cite{kim_book}.
By integrating Eq.\ \eqref{eq:hismoluchowski} with
$N\int \mathrm{d}{\mathbf r}_{2}\;...\int\mathrm{d}\mathbf{r}_{N}$,
we obtain the equation for the time-evolution of the one-body density
$\rho(\mathbf{r},t)$.
The resulting equation depends on both, the time-dependent two-body
and the three-body densities.
We cast those into a form involving exclusively the equilibrium
Helmholtz free energy functional 
$\mathcal{F}[\rho]= k_{\mathrm{B}}T
\int \mathrm{d}\mathbf{r}\; \rho(\mathbf{r},t)[
\ln(\Lambda^{3}\rho(\mathbf{r},t))-1]+
\mathcal{F}_{\mathrm{exc}}[\rho] +\int \mathrm{d}\mathbf{r}\;
\rho(\mathbf{r},t) V_{\mathrm{ext}}(\mathbf{r},t)$, with
$\mathcal{F}_{\mathrm{exc}}[\rho]$ being the excess contribution to
the free energy functional and $\Lambda$ the thermal de Broglie
wavelength, by making use of static DFT \cite{evans_ap_1979} and the
first two members of the Yvon-Born-Green (YBG) relations (see, e.g.,
\cite{hansen06}).
%DFT is a very powerful tool for the quantitative description of
%the equilibrium states of many-body systems under arbitrary static
%external fields.
%It is based on the exact statement that for a given interparticle
%interaction the intrinsic free energy functional 
%$\mathcal{F}[\rho]$ is a unique functional of the inhomogeneous
%one-particle density.
To that end, we identify the out-of-equilibrium system at each
point in time with an equilibrium reference system whose density
profiles are identical.
The basic assumption now, which also underlies the original
version of the DDFT
\cite{tarazona_jpcm_2000,tarazona_jcp_1999,archer_jcp_2004}, is to 
approximate the nonequilibrium two-body and three-body densities by
those of the reference system with the same one-body density.
Thus, we obtain our central result:
\begin{eqnarray}
  \label{eq:hiddft}
  \nonumber
  \Gamma^{-1}\frac{\ensuremath{\partial{\rho(\mathbf{r},t)}}}
  {\ensuremath{\partial{t}}} = 
  \mathbf{\nabla}_{\mathbf{r}}\cdot\Bigg\{
  \rho(\mathbf{r},t)\mathbf{\nabla}_{\mathbf{r}} \frac{\delta
  \mathcal{F}[\rho]}{\delta \rho(\mathbf{r},t)}+\\
  \nonumber
  \int \mathrm{d}{\mathbf r}^{\prime}\;
  \rho^{(2)}(\mathbf{r},\mathbf{r}^{\prime},t)
  \bm{\omega}_{11}(\mathbf{r}-\mathbf{r}^{\prime})
  \cdot\mathbf{\nabla}_{\mathbf{r}} \;
  \frac{\delta \mathcal{F}[\rho]}{\delta \rho(\mathbf{r},t)}+\\
  \int \mathrm{d}{\mathbf r}^{\prime}\;
  \rho^{(2)}(\mathbf{r},\mathbf{r}^{\prime},t)
  \bm{\omega}_{12}(\mathbf{r}-\mathbf{r}^{\prime})
  \cdot\mathbf{\nabla}_{\mathbf{r}^{\prime}}
  \frac{\delta \mathcal{F}[\rho]}{\delta
  \rho(\mathbf{r}^{\prime},t)}\Bigg \},
\end{eqnarray}
with the mobility constant $\Gamma$ for which the Einstein relation
gives $D_{0}/\Gamma=(k_{\mathrm{B}}T)^{-1}$.
Eq.\ \eqref{eq:hiddft} has the form of a continuity equation with the
current density $\mathbf{j}=\mathbf{j}_{1} + \mathbf{j}_{2} +
\mathbf{j}_{3}$ given by the terms in the curly brackets.
The current density $\mathbf{j}_{1}$ is proportional to the thermodynamic
force $\mathbf{\nabla}_{\mathbf{r}}\frac{\delta
\mathcal{F}[\rho]}{\delta \rho(\mathbf{r},t)}$ and 
 persists when hydrodynamic interactions are
neglected \cite{tarazona_jpcm_2000,tarazona_jcp_1999}.
$\mathbf{j}_{2}$ and $\mathbf{j}_{3}$ are additional current densities
which occur due to the solvent mediated hydrodynamic interactions.
$\mathbf{j}_{2}$ describes the current density stemming from the
reflection of the solvent flow induced by the thermodynamic
force at position $\mathbf{r}$ on the surrounding particles.
$\mathbf{j}_{3}$, on the other hand, is the current density at
position $\mathbf{r}$ due to the solvent flow induced by the
thermodynamic force at position $\mathbf{r}^{\prime}$.

Finally, we close the above relation Eq.\ \eqref{eq:hiddft},
which still depends on the nonequilibrium two-body density
$\rho^{(2)}(\mathbf{r},\mathbf{r}^{\prime},t)$.
Within our approximation -- the two-body density is assumed to be
identical to the equilibrium one of the reference system -- it is
given at every point in time by the exact generalized Ornstein-Zernike
equation \cite{evans_ap_1979}:
\begin{eqnarray}
  \label{eq:hioz}
  \nonumber
  &\rho^{(2)}(\mathbf{r},\mathbf{r}^{\prime},t) =
  \rho(\mathbf{r},t) \rho(\mathbf{r}^{\prime},t)
  \left(1+(k_{\mathrm{B}}T)^{-1}
  \frac{\delta^{2}\mathcal{F}_{\mathrm{exc}}[\rho]}{\delta
  \rho(\mathbf{r},t) \delta \rho(\mathbf{r}^{\prime},t) }\right) +\\
  \nonumber
  &\rho(\mathbf{r}^{\prime},t) \int \mathrm{d}{\mathbf
  r}^{\prime\prime}\;\bigg\{
  (\rho^{(2)}(\mathbf{r},\mathbf{r}^{\prime\prime},t)-
  \rho(\mathbf{r},t) \rho(\mathbf{r}^{\prime\prime},t))\\
  &(k_{\mathrm{B}}T)^{-1} \frac{\delta^{2}\mathcal{F}_{\mathrm{exc}}[\rho]}{\delta
  \rho(\mathbf{r}^{\prime},t) \delta
  \rho(\mathbf{r}^{\prime\prime},t)}\bigg\}.
\end{eqnarray}
This implicit equation for the two-body density of the inhomogeneous
system may be reasonably approximated by its bulk value 
\cite{haase_pr_1995,goetzelmann_pre_1996}, i.e.:
$\rho^{(2)}(\mathbf{r},\mathbf{r}^{\prime},t)\approx\rho(r,t)
\rho(r^{\prime},t)g(|\mathbf{r}-\mathbf{r}^{\prime}|,{\bar \rho})$, where
$g(|\mathbf{r}-\mathbf{r}^{\prime}|, {\bar \rho})$ is the pair correlation function
for a homogeneous system at an appropriately averaged density ${\bar \rho}$.
For a hard sphere fluid, an analytic expression for the pair
correlation function is available based on the Percus-Yevick equation
\cite{trokhymchuk_jcp_2005}.

\begin{figure}[th]
  \begin{center}
    \includegraphics[width=8cm, clip=true, draft=false]{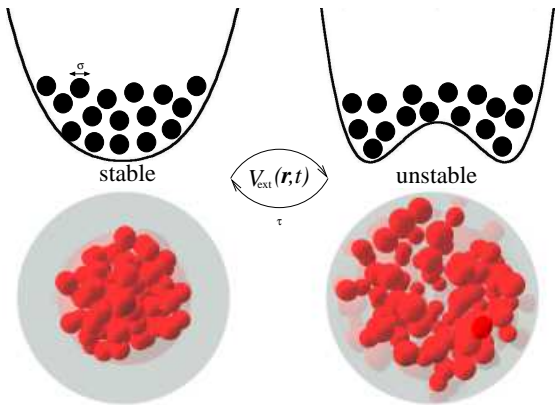}
    \caption{(Color online) Sketch of the confined system.
    The external potential models an optical trap $V_{ext}(r,t)$ which
    changes its central shape from stable to unstable within a time
    period $\tau$.
    The trap confines $N$ colloidal hard spheres of diameter $\sigma$
    shown as black circles.
    Additionally, typical 3d simulation snapshots are shown.
    The left hand side shows an initial stable configuration for $t=0$ and
    the right hand side shows an unstable situation at
    $t=2.75\tau_{\mathrm{B}}$ for case (N) \cite{prlsupplement_ddfthi}.}
    \label{fig:sketch}
  \end{center}
\end{figure}

We use the method presented here to investigate the time-evolution of
the one-body density of a confined cluster of $N=100$ monodisperse hard
spherical particles of diameter $\sigma$, which serves as the unit of
length henceforth.
An appropriate time scale is $\tau_\mathrm{B}=\sigma^{2}/D_{0}$, and
the energy unit is $k_\mathrm{B} T$.
The particles are trapped in a 
 soft spherical cavity which switches from a stable to an unstable
shape periodically in time. The confining external potential
 only acts on the colloidal particles. Therefore
 the solvent is treated as an unbounded fluid.
The total external potential is modeled as
\begin{equation}
  \label{eq:extpotcav}
  V_{\mathrm{ext}}(r,t)=V_{1}\left(\frac{r}{R_{1}}\right)^{4}+
  V_{2} \cos(2\pi t/\tau) \left(\frac{r}{R_{2}}\right)^{2},
\end{equation}
where $r=|\mathbf{r}|$, $R_{1}=4\sigma$ and $V_{1}=10k_\mathrm{B} T$
are the length scale and the strength of an outer fixed cavity and
$R_{2}=\sigma$ and $V_{2}=k_\mathrm{B} T$ are the
length scale and strength of an inner part, which oscillates in time
with a period $\tau=0.5\tau_\mathrm{B}$.
A sketch of the setup is shown in Fig.\ \ref{fig:sketch}.
Due to the spherical symmetry, the density profile $\rho(r,t)$ depends
only on the radial position coordinate $r$.

For the hard sphere excess density functional $\mathcal{F}_{\mathrm{exc}}[\rho]$
Rosenfeld's fundamental measure theory \cite{rosenfeld_prl_1989} was used
which provides a very reliable approximation in equilibrium
\cite{gonzalez_prl_1997}.
The distinct hydrodynamic tensor $\bm{\omega}_{12}(\mathbf{r})$ is 
approximated by the Rotne-Prager
expression \cite{rotne_jcp_1969}
$D_{0}\bm{\omega}_{12}(\mathbf{r})=\frac{3}{8}(\frac{\sigma}{r})
[\mathbf{1}+\hat{\mathbf{r}}\hat{\mathbf{r}}] +
\frac{1}{16}(\frac{\sigma}{r})^{3}
[\mathbf{1}-3\hat{\mathbf{r}}\hat{\mathbf{r}}]
+\mathcal{O}[(\frac{\sigma}{r})^{7}]$
while the self term $\bm{\omega}_{11}(\mathbf{r})$ whose leading order term
is $\mathcal{O}((\sigma/r)^{4})$ is neglected.
Here, $\hat{\mathbf{r}}=\mathbf{r}/|\mathbf{r}|$ denotes the unit vector,
$\hat{\mathbf{r}}\hat{\mathbf{r}}$ is the dyadic product, and
$\mathbf{1}$ is the unit matrix. 
Thus, on this level of approximation, we incorporate all solvent
mediated interactions up to order of $\mathcal{O}((\sigma/r)^{3})$.
The pair correlation $g(|\mathbf{r}-\mathbf{r}^{\prime}|, {\bar \rho})$
is calculated at each time step at the average density of the system
$\bar{\rho}(t)=1/R_{\mathrm{max}}(t)\int_{0}^{R_{\mathrm{max}}(t)}
\mathrm{d}r \rho(r,t)$, where $R_{\mathrm{max}}$ is defined by
$V_{\mathrm{ext}}(r=R_{\mathrm{max}}(t))=10k_{\mathrm{B}}T$.

The results are tested against Brownian dynamics simulations
\cite{allen_tildesley_book} performed on the same level of accuracy of
the diffusion tensor, in which the hard interaction is approximated by
a slightly softened one:
\begin{equation}
  \label{eq:softcore_potential}
  \frac{v_{2}(r)}{k_\mathrm{B} T} = \left\{
  \begin{array}{cl}
    \left [ \left (
      \frac{\sigma}{r}\right)^{48} - \left (
    \frac{\sigma}{r}\right)^{24}
    +\frac{1}{4}\right] &
  \textrm{if $r \leq
  2^{1/24}\sigma$}\\
  0 & \textrm{else}
  \end{array}
  \right. .
\end{equation}
In all simulations we chose a finite simulation time step of $\Delta
t=10^{-4}\tau_{\mathrm{B}}$.
In order to obtain the time-dependent density $\rho(r,t)$ we
perform a large number of $N_{\mathrm{run}}=10^{4}$ independent runs
with different initial configurations sampled from a situation with a
static external potential, i.e., Eq.\ \eqref{eq:extpotcav} at $t=0$.
Additionally, the densities are compared to those obtained by standard
DDFT where hydrodynamic interactions are ignored, i.e.,
$\bm{\omega}_{11}=\bm{\omega}_{12}= 0$.
Henceforth, we label the situation including hydrodynamic interactions
(H) and the situation where they are neglected (N), respectively.
The initial density profile is the equilibrium density profile
for $V_{\mathrm{ext}}(r,t=0)$ calculated from static density
functional theory.
Typical simulation snapshots are shown in Fig.\ \ref{fig:sketch}.

\begin{figure}[ht]
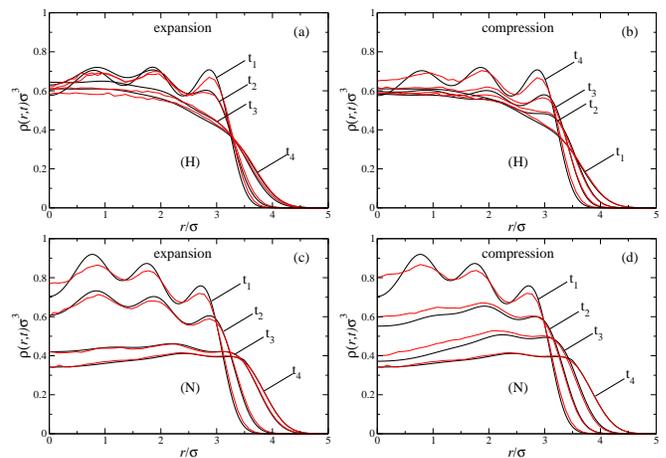

  \begin{center}
    \includegraphics[width=4.25cm,clip=true,draft=false]{fig2a}
    \includegraphics[width=4.25cm,clip=true,draft=false]{fig2b}
    \includegraphics[width=4.25cm,clip=true,draft=false]{fig2c}
    \includegraphics[width=4.25cm,clip=true,draft=false]{fig2d}
    \caption{(Color online) {\it Steady-state} DDFT (solid curves) and
    BD (noisy curves) results for the time dependent density profile
    $\rho(r,t)$. In Fig.\ (a) and (b) hydrodynamic interactions are
    taken into account while in (c) and (d) they are neglected.
    (a) and (c) correspond to the expanding half period and (b) and
    (d) to the compressing half period, respectively
    \cite{prlsupplement_ddfthi}.
    The profiles correspond to the following time sequence:
    $t_{0}=2.5\tau_{\mathrm{B}}$,
    $t_{1}=2.6\tau_{\mathrm{B}}$, 
    $t_{2}=2.7\tau_{\mathrm{B}}$, 
    $t_{3}=2.75\tau_{\mathrm{B}}$ in (a) and (c), and 
    $t_{4}=2.75\tau_{\mathrm{B}}$, 
    $t_{5}=2.85\tau_{\mathrm{B}}$, 
    $t_{5}=2.9\tau_{\mathrm{B}}$, and
    $t_{6}=3.0\tau_{\mathrm{B}}$.}
    \label{fig:profiles}
  \end{center}
\end{figure}
The resulting steady-state of the dynamical density profiles, i.e.\ the
driven breathing mode after initial relaxation, is depicted in
 Fig.\ \ref{fig:profiles}. First of all, theory and simulation results
are in very good agreement for both situations (H) and (N)
but we observe distinct qualitative differences in the breathing mode:
Hydrodynamic interactions tend to damp the density response
considerably.
For neglected hydrodynamic interactions there are huge density
oscillations, in particular at the trap center but also at the trap
boundaries which are significantly smaller for hydrodynamic
interactions.
This result is not obvious as hydrodynamic interactions tend to
accelerate neighboring particle which are driven into the same
direction.
The damping effects seen here is caused by the overall motion of the
breathing mode which hinders collective streaming due to the counter 
motion in the opposed part of the trap.

\begin{figure}[ht]
  \begin{center}
    \includegraphics[width=8.0cm,clip=true,draft=false]{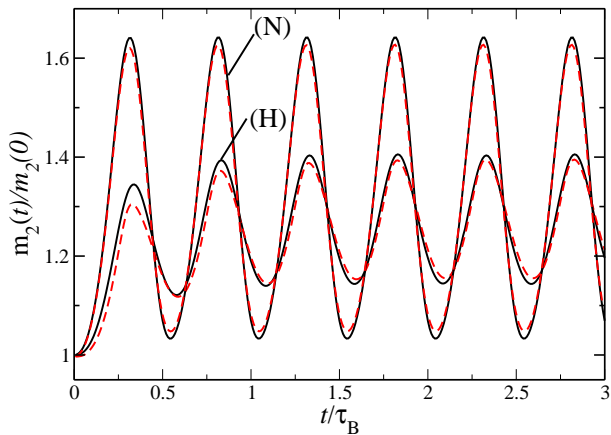}
    \caption{(Color online) Second moment of the breathing mode,
    $m_{2}(t)$, versus time $t$. DDFT (solid curves) and
    BD (dashed curves) results for hydrodynamic interactions taken
    into account (H) and being neglected (N).}
    \label{fig:moment}
  \end{center}
\end{figure}

In order to analyze the relaxational behavior towards the steady
state, we introduce the second moment of the breathing mode
$m_{2}(t)=\int\mathrm{d}\mathbf{r}\;r^{2}\rho(r,t)$.
It is shown in Fig.\ \ref{fig:moment} in both cases (H) and (N) for
DDFT and Brownian Dynamics.
Clearly, the dynamic evolution of the second moment is strongly
damped by hydrodynamic interactions as revealed by the much slower
oscillation amplitude.
On top of that the relaxation time towards the steady state breathing
mode is considerably larger for hydrodynamic interactions as compared
to the simple Brownian case where the relaxation is almost
instantaneous.
The second moment of the breathing mode is slightly off-phase with
respect to the driving external potential $\sim \cos(2\pi t/\tau)$, and
the hydrodynamic interactions lead to a stronger phase-shifting.

In conclusion, we have proposed a dynamical density functional theory
which includes hydrodynamic interactions between the colloidal
particles and applied it to access the driven breathing mode in
oscillating optical traps.
The theory was confirmed by Brownian dynamics computer simulations.
Hydrodynamic interactions were found to damp the response to the
driving trap, to increase the relaxation time towards the steady state
and to increase the phase shift.
These predictions can in principle be tested by real-space experiments
on confined colloidal particles.
For charged suspensions, the strength of the hydrodynamic interactions
can be systematically tuned in the experiments by varying the colloid
charge which governs size of the interactions relative to the physical
core which governs the hydrodynamic interactions.
It would be interesting to generalize the theory further to
orientational degrees of freedom and to self-propelled colloidal
particles representing microswimmers where hydrodynamic interactions
play a key role.

\begin{acknowledgments}
We thank G.\ N\"agele, S.\ van Teeffelen, and P.\ Royall for helpful
discussions.
This work is supported by the DFG within SFB TR6 (project section D3)
and by the Graduiertenf\"orderung of the University of D\"usseldorf.
\end{acknowledgments}

\bibliography{/home/rexm/Documents/bibliography_pre}

\begin{thebibliography}{35}
\expandafter\ifx\csname natexlab\endcsname\relax\def\natexlab#1{#1}\fi
\expandafter\ifx\csname bibnamefont\endcsname\relax
  \def\bibnamefont#1{#1}\fi
\expandafter\ifx\csname bibfnamefont\endcsname\relax
  \def\bibfnamefont#1{#1}\fi
\expandafter\ifx\csname citenamefont\endcsname\relax
  \def\citenamefont#1{#1}\fi
\expandafter\ifx\csname url\endcsname\relax
  \def\url#1{\texttt{#1}}\fi
\expandafter\ifx\csname urlprefix\endcsname\relax\def\urlprefix{URL }\fi
\providecommand{\bibinfo}[2]{#2}
\providecommand{\eprint}[2][]{\url{#2}}

\bibitem[{\citenamefont{Ma et~al.}(2003)\citenamefont{Ma, Wen, Tam, and
  Sheng}}]{ma_ap_2003}
\bibinfo{author}{\bibfnamefont{H.~R.} \bibnamefont{Ma}},
  \bibinfo{author}{\bibfnamefont{W.~J.} \bibnamefont{Wen}},
  \bibinfo{author}{\bibfnamefont{W.~Y.} \bibnamefont{Tam}}, \bibnamefont{and}
  \bibinfo{author}{\bibfnamefont{P.}~\bibnamefont{Sheng}},
  \bibinfo{journal}{Adv.\ in Phys.} \textbf{\bibinfo{volume}{52}},
  \bibinfo{pages}{343} (\bibinfo{year}{2003}).

\bibitem[{\citenamefont{Sciortino and Tartaglia}(2005)}]{sciortino_ap_2005}
\bibinfo{author}{\bibfnamefont{F.}~\bibnamefont{Sciortino}} \bibnamefont{and}
  \bibinfo{author}{\bibfnamefont{P.}~\bibnamefont{Tartaglia}},
  \bibinfo{journal}{Adv.\ in Phys.} \textbf{\bibinfo{volume}{54}},
  \bibinfo{pages}{471} (\bibinfo{year}{2005}).

\bibitem[{\citenamefont{Lutz et~al.}(2006)\citenamefont{Lutz, Reichert, Stark,
  and Bechinger}}]{lutz_epl_2006}
\bibinfo{author}{\bibfnamefont{C.}~\bibnamefont{Lutz}},
  \bibinfo{author}{\bibfnamefont{M.}~\bibnamefont{Reichert}},
  \bibinfo{author}{\bibfnamefont{H.}~\bibnamefont{Stark}}, \bibnamefont{and}
  \bibinfo{author}{\bibfnamefont{C.}~\bibnamefont{Bechinger}},
  \bibinfo{journal}{Europhys.\ Lett.} \textbf{\bibinfo{volume}{74}},
  \bibinfo{pages}{719} (\bibinfo{year}{2006}).

\bibitem[{\citenamefont{Banchio et~al.}(2006)\citenamefont{Banchio, Gapinski,
  Patkowski, Hauszler, Fluerasu, Sacanna, Holmqvist, Meier, Lettinga, and
  N{\"a}gele}}]{banchio_prl_2006}
\bibinfo{author}{\bibfnamefont{A.}~\bibnamefont{Banchio}},
  \bibinfo{author}{\bibfnamefont{J.}~\bibnamefont{Gapinski}},
  \bibinfo{author}{\bibfnamefont{A.}~\bibnamefont{Patkowski}},
  \bibinfo{author}{\bibfnamefont{W.}~\bibnamefont{Hauszler}},
  \bibinfo{author}{\bibfnamefont{A.}~\bibnamefont{Fluerasu}},
  \bibinfo{author}{\bibfnamefont{S.}~\bibnamefont{Sacanna}},
  \bibinfo{author}{\bibfnamefont{P.}~\bibnamefont{Holmqvist}},
  \bibinfo{author}{\bibfnamefont{G.}~\bibnamefont{Meier}},
  \bibinfo{author}{\bibfnamefont{M.}~\bibnamefont{Lettinga}}, \bibnamefont{and}
  \bibinfo{author}{\bibfnamefont{G.}~\bibnamefont{N{\"a}gele}},
  \bibinfo{journal}{Phys.\ Rev.\ Lett.} \textbf{\bibinfo{volume}{96}},
  \bibinfo{pages}{138303} (\bibinfo{year}{2006}).

\bibitem[{\citenamefont{Padding and Louis}(2004)}]{padding_prl_2004}
\bibinfo{author}{\bibfnamefont{J.~T.} \bibnamefont{Padding}} \bibnamefont{and}
  \bibinfo{author}{\bibfnamefont{A.~A.} \bibnamefont{Louis}},
  \bibinfo{journal}{Phys.\ Rev.\ Lett.} \textbf{\bibinfo{volume}{93}},
  \bibinfo{pages}{220601} (\bibinfo{year}{2004}).

\bibitem[{\citenamefont{Kuusela et~al.}(2003)\citenamefont{Kuusela, Lahtinen,
  and Ala-Nissila}}]{kuusela_prl_2003}
\bibinfo{author}{\bibfnamefont{E.}~\bibnamefont{Kuusela}},
  \bibinfo{author}{\bibfnamefont{J.~M.} \bibnamefont{Lahtinen}},
  \bibnamefont{and}
  \bibinfo{author}{\bibfnamefont{T.}~\bibnamefont{Ala-Nissila}},
  \bibinfo{journal}{Phys.\ Rev.\ Lett.} \textbf{\bibinfo{volume}{90}},
  \bibinfo{pages}{094502} (\bibinfo{year}{2003}).

\bibitem[{\citenamefont{N{\"a}gele and Dhont}(1998)}]{naegele_jcp_1998_a}
\bibinfo{author}{\bibfnamefont{G.}~\bibnamefont{N{\"a}gele}} \bibnamefont{and}
  \bibinfo{author}{\bibfnamefont{J.~K.~G.} \bibnamefont{Dhont}},
  \bibinfo{journal}{J.\ Chem.\ Phys.} \textbf{\bibinfo{volume}{108}},
  \bibinfo{pages}{9566} (\bibinfo{year}{1998}).

\bibitem[{\citenamefont{N{\"a}gele and Bergenholtz}(1998)}]{naegele_jcp_1998_b}
\bibinfo{author}{\bibfnamefont{G.}~\bibnamefont{N{\"a}gele}} \bibnamefont{and}
  \bibinfo{author}{\bibfnamefont{J.}~\bibnamefont{Bergenholtz}},
  \bibinfo{journal}{J.\ Chem.\ Phys.} \textbf{\bibinfo{volume}{108}},
  \bibinfo{pages}{9893} (\bibinfo{year}{1998}).

\bibitem[{\citenamefont{N{\"a}gele et~al.}(1999)\citenamefont{N{\"a}gele,
  Bergenholtz, and Dhont}}]{naegele_jcp_1999}
\bibinfo{author}{\bibfnamefont{G.}~\bibnamefont{N{\"a}gele}},
  \bibinfo{author}{\bibfnamefont{J.}~\bibnamefont{Bergenholtz}},
  \bibnamefont{and} \bibinfo{author}{\bibfnamefont{J.~K.~G.}
  \bibnamefont{Dhont}}, \bibinfo{journal}{J.\ Chem.\ Phys.}
  \textbf{\bibinfo{volume}{110}}, \bibinfo{pages}{7037} (\bibinfo{year}{1999}).

\bibitem[{\citenamefont{Cichocki et~al.}(2002)\citenamefont{Cichocki,
  Ekiel-Jezewska, Szymczak, and Wajnryb}}]{cichocki_jcp_2002}
\bibinfo{author}{\bibfnamefont{B.}~\bibnamefont{Cichocki}},
  \bibinfo{author}{\bibfnamefont{M.~L.} \bibnamefont{Ekiel-Jezewska}},
  \bibinfo{author}{\bibfnamefont{P.}~\bibnamefont{Szymczak}}, \bibnamefont{and}
  \bibinfo{author}{\bibfnamefont{E.}~\bibnamefont{Wajnryb}},
  \bibinfo{journal}{J.\ Chem.\ Phys.} \textbf{\bibinfo{volume}{117}},
  \bibinfo{pages}{1231} (\bibinfo{year}{2002}).

\bibitem[{\citenamefont{Dhont}(1996)}]{dhont_book}
\bibinfo{author}{\bibfnamefont{J.~K.~G.} \bibnamefont{Dhont}},
  \emph{\bibinfo{title}{An Introduction to Dynamics of Colloids}}
  (\bibinfo{publisher}{Elsevier Science}, \bibinfo{address}{Amsterdam},
  \bibinfo{year}{1996}).

\bibitem[{\citenamefont{N{\"a}gele}(2003)}]{naegele_jpcm_2003}
\bibinfo{author}{\bibfnamefont{G.}~\bibnamefont{N{\"a}gele}},
  \bibinfo{journal}{J.\ Phys.: Condens.\ Matter} \textbf{\bibinfo{volume}{15}},
  \bibinfo{pages}{S407} (\bibinfo{year}{2003}).

\bibitem[{\citenamefont{Pesch\'e and N{\"a}gele}(2000)}]{pesche_epl_2000}
\bibinfo{author}{\bibfnamefont{R.}~\bibnamefont{Pesch\'e}} \bibnamefont{and}
  \bibinfo{author}{\bibfnamefont{G.}~\bibnamefont{N{\"a}gele}},
  \bibinfo{journal}{Europhys.\ Lett.} \textbf{\bibinfo{volume}{51}},
  \bibinfo{pages}{584} (\bibinfo{year}{2000}).

\bibitem[{\citenamefont{Lionberger and Russel}(1997)}]{lionberger_jr_1997}
\bibinfo{author}{\bibfnamefont{R.~A.} \bibnamefont{Lionberger}}
  \bibnamefont{and} \bibinfo{author}{\bibfnamefont{W.~B.}
  \bibnamefont{Russel}}, \bibinfo{journal}{J.\ Rheol.}
  \textbf{\bibinfo{volume}{41}}, \bibinfo{pages}{399} (\bibinfo{year}{1997}).

\bibitem[{\citenamefont{Chan and Finken}(2005)}]{chan_prl_2005}
\bibinfo{author}{\bibfnamefont{G.~K.~L.} \bibnamefont{Chan}} \bibnamefont{and}
  \bibinfo{author}{\bibfnamefont{R.}~\bibnamefont{Finken}},
  \bibinfo{journal}{Phys.\ Rev.\ Lett.} \textbf{\bibinfo{volume}{94}},
  \bibinfo{pages}{183001} (\bibinfo{year}{2005}).

\bibitem[{\citenamefont{Archer and Evans}(2004)}]{archer_jcp_2004}
\bibinfo{author}{\bibfnamefont{A.~J.} \bibnamefont{Archer}} \bibnamefont{and}
  \bibinfo{author}{\bibfnamefont{R.}~\bibnamefont{Evans}},
  \bibinfo{journal}{J.\ Chem.\ Phys.} \textbf{\bibinfo{volume}{121}},
  \bibinfo{pages}{4246} (\bibinfo{year}{2004}).

\bibitem[{\citenamefont{Marconi and Tarazona}(1999)}]{tarazona_jcp_1999}
\bibinfo{author}{\bibfnamefont{U.~M.~B.} \bibnamefont{Marconi}}
  \bibnamefont{and} \bibinfo{author}{\bibfnamefont{P.}~\bibnamefont{Tarazona}},
  \bibinfo{journal}{J.\ Chem.\ Phys.} \textbf{\bibinfo{volume}{110}},
  \bibinfo{pages}{8032} (\bibinfo{year}{1999}).

\bibitem[{\citenamefont{Marconi and Tarazona}(2000)}]{tarazona_jpcm_2000}
\bibinfo{author}{\bibfnamefont{U.~M.~B.} \bibnamefont{Marconi}}
  \bibnamefont{and} \bibinfo{author}{\bibfnamefont{P.}~\bibnamefont{Tarazona}},
  \bibinfo{journal}{J.\ Phys.: Condens.\ Matter} \textbf{\bibinfo{volume}{12}},
  \bibinfo{pages}{A413} (\bibinfo{year}{2000}).

\bibitem[{\citenamefont{Henderson et~al.}(2002)\citenamefont{Henderson,
  Mitchell, and Bartlett}}]{henderson_prl_2002}
\bibinfo{author}{\bibfnamefont{S.}~\bibnamefont{Henderson}},
  \bibinfo{author}{\bibfnamefont{S.}~\bibnamefont{Mitchell}}, \bibnamefont{and}
  \bibinfo{author}{\bibfnamefont{P.}~\bibnamefont{Bartlett}},
  \bibinfo{journal}{Phys.\ Rev.\ Lett.} \textbf{\bibinfo{volume}{88}},
  \bibinfo{pages}{088302} (\bibinfo{year}{2002}).

\bibitem[{\citenamefont{Martin et~al.}(2006)\citenamefont{Martin, Reichert,
  Stark, and Gisler}}]{martin_prl_2006}
\bibinfo{author}{\bibfnamefont{S.}~\bibnamefont{Martin}},
  \bibinfo{author}{\bibfnamefont{M.}~\bibnamefont{Reichert}},
  \bibinfo{author}{\bibfnamefont{H.}~\bibnamefont{Stark}}, \bibnamefont{and}
  \bibinfo{author}{\bibfnamefont{T.}~\bibnamefont{Gisler}},
  \bibinfo{journal}{Phys.\ Rev.\ Lett.} \textbf{\bibinfo{volume}{97}},
  \bibinfo{pages}{248301} (\bibinfo{year}{2006}).

\bibitem[{\citenamefont{Griffin et~al.}(1997)\citenamefont{Griffin, Wu, and
  Stringari}}]{griffin_prl_1997}
\bibinfo{author}{\bibfnamefont{A.}~\bibnamefont{Griffin}},
  \bibinfo{author}{\bibfnamefont{W.~C.} \bibnamefont{Wu}}, \bibnamefont{and}
  \bibinfo{author}{\bibfnamefont{S.}~\bibnamefont{Stringari}},
  \bibinfo{journal}{Phys.\ Rev.\ Lett.} \textbf{\bibinfo{volume}{78}},
  \bibinfo{pages}{1838} (\bibinfo{year}{1997}).

\bibitem[{\citenamefont{Guidarelli et~al.}(1998)\citenamefont{Guidarelli,
  Craciun, Galassi, and Roncari}}]{guidarelli_us_1998}
\bibinfo{author}{\bibfnamefont{G.}~\bibnamefont{Guidarelli}},
  \bibinfo{author}{\bibfnamefont{F.}~\bibnamefont{Craciun}},
  \bibinfo{author}{\bibfnamefont{C.}~\bibnamefont{Galassi}}, \bibnamefont{and}
  \bibinfo{author}{\bibfnamefont{E.}~\bibnamefont{Roncari}},
  \bibinfo{journal}{Ultrasonics} \textbf{\bibinfo{volume}{36}},
  \bibinfo{pages}{467} (\bibinfo{year}{1998}).

\bibitem[{\citenamefont{Chowdhury et~al.}(1985)\citenamefont{Chowdhury,
  Ackerson, and Clark}}]{chowdhury_prl_1985}
\bibinfo{author}{\bibfnamefont{A.}~\bibnamefont{Chowdhury}},
  \bibinfo{author}{\bibfnamefont{B.~J.} \bibnamefont{Ackerson}},
  \bibnamefont{and} \bibinfo{author}{\bibfnamefont{N.~A.} \bibnamefont{Clark}},
  \bibinfo{journal}{Phys.\ Rev.\ Lett.} \textbf{\bibinfo{volume}{55}},
  \bibinfo{pages}{833} (\bibinfo{year}{1985}).

\bibitem[{\citenamefont{Archer}(2006)}]{archer_jpcm_2006}
\bibinfo{author}{\bibfnamefont{A.~J.} \bibnamefont{Archer}},
  \bibinfo{journal}{J.\ Phys.: Condens.\ Matter} \textbf{\bibinfo{volume}{18}},
  \bibinfo{pages}{5617} (\bibinfo{year}{2006}).

\bibitem[{\citenamefont{Kim and Karrila}(1991)}]{kim_book}
\bibinfo{author}{\bibfnamefont{S.}~\bibnamefont{Kim}} \bibnamefont{and}
  \bibinfo{author}{\bibfnamefont{S.~J.} \bibnamefont{Karrila}},
  \emph{\bibinfo{title}{Microhydrodynamics: Principles and selected
  applications}} (\bibinfo{publisher}{Butterworth-Heinemann},
  \bibinfo{address}{Boston}, \bibinfo{year}{1991}).

\bibitem[{\citenamefont{Evans}(1979)}]{evans_ap_1979}
\bibinfo{author}{\bibfnamefont{R.}~\bibnamefont{Evans}},
  \bibinfo{journal}{Adv.\ in Phys.} \textbf{\bibinfo{volume}{28}},
  \bibinfo{pages}{143} (\bibinfo{year}{1979}).

\bibitem[{\citenamefont{Hansen and MacDonald}(2006)}]{hansen06}
\bibinfo{author}{\bibfnamefont{J.~P.} \bibnamefont{Hansen}} \bibnamefont{and}
  \bibinfo{author}{\bibfnamefont{I.~R.} \bibnamefont{MacDonald}},
  \emph{\bibinfo{title}{Theory of Simple Liquids}}
  (\bibinfo{publisher}{Academic}, \bibinfo{address}{London},
  \bibinfo{year}{2006}), \bibinfo{edition}{3rd} ed.

\bibitem[{\citenamefont{Dietrich and Haase}(1995)}]{haase_pr_1995}
\bibinfo{author}{\bibfnamefont{S.}~\bibnamefont{Dietrich}} \bibnamefont{and}
  \bibinfo{author}{\bibfnamefont{A.}~\bibnamefont{Haase}},
  \bibinfo{journal}{Phys.\ Rep.} \textbf{\bibinfo{volume}{260}},
  \bibinfo{pages}{1} (\bibinfo{year}{1995}).

\bibitem[{\citenamefont{G{\"o}tzelmann
  et~al.}(1996)\citenamefont{G{\"o}tzelmann, Haase, and
  Dietrich}}]{goetzelmann_pre_1996}
\bibinfo{author}{\bibfnamefont{B.}~\bibnamefont{G{\"o}tzelmann}},
  \bibinfo{author}{\bibfnamefont{A.}~\bibnamefont{Haase}}, \bibnamefont{and}
  \bibinfo{author}{\bibfnamefont{S.}~\bibnamefont{Dietrich}},
  \bibinfo{journal}{Phys.\ Rev.\ E} \textbf{\bibinfo{volume}{53}},
  \bibinfo{pages}{3456} (\bibinfo{year}{1996}).

\bibitem[{\citenamefont{Trokhymchuk et~al.}(2005)\citenamefont{Trokhymchuk,
  Nezbeda, Jirs{\'a}k, and Henderson}}]{trokhymchuk_jcp_2005}
\bibinfo{author}{\bibfnamefont{A.}~\bibnamefont{Trokhymchuk}},
  \bibinfo{author}{\bibfnamefont{I.}~\bibnamefont{Nezbeda}},
  \bibinfo{author}{\bibfnamefont{J.}~\bibnamefont{Jirs{\'a}k}},
  \bibnamefont{and}
  \bibinfo{author}{\bibfnamefont{D.}~\bibnamefont{Henderson}},
  \bibinfo{journal}{J.\ Chem.\ Phys.} \textbf{\bibinfo{volume}{123}},
  \bibinfo{pages}{024501} (\bibinfo{year}{2005}).

\bibitem[{prl()}]{prlsupplement_ddfthi}
\bibinfo{note}{See EPAPS Document No. [] for movies of the time evolution of
  $\rho({\bf r},t)$ and typical simulation runs for both setups (H) and (N).}

\bibitem[{\citenamefont{Rosenfeld}(1989)}]{rosenfeld_prl_1989}
\bibinfo{author}{\bibfnamefont{Y.}~\bibnamefont{Rosenfeld}},
  \bibinfo{journal}{Phys.\ Rev.\ Lett.} \textbf{\bibinfo{volume}{63}},
  \bibinfo{pages}{980} (\bibinfo{year}{1989}).

\bibitem[{\citenamefont{Gonz{\'a}lez et~al.}(1997)\citenamefont{Gonz{\'a}lez,
  White, Rom{\'a}n, Velasco, and Evans}}]{gonzalez_prl_1997}
\bibinfo{author}{\bibfnamefont{A.}~\bibnamefont{Gonz{\'a}lez}},
  \bibinfo{author}{\bibfnamefont{J.~A.} \bibnamefont{White}},
  \bibinfo{author}{\bibfnamefont{F.~L.} \bibnamefont{Rom{\'a}n}},
  \bibinfo{author}{\bibfnamefont{S.}~\bibnamefont{Velasco}}, \bibnamefont{and}
  \bibinfo{author}{\bibfnamefont{R.}~\bibnamefont{Evans}},
  \bibinfo{journal}{Phys.\ Rev.\ Lett.} \textbf{\bibinfo{volume}{79}},
  \bibinfo{pages}{2466} (\bibinfo{year}{1997}).

\bibitem[{\citenamefont{Rotne and Prager}(1969)}]{rotne_jcp_1969}
\bibinfo{author}{\bibfnamefont{J.}~\bibnamefont{Rotne}} \bibnamefont{and}
  \bibinfo{author}{\bibfnamefont{S.}~\bibnamefont{Prager}},
  \bibinfo{journal}{J.\ Chem.\ Phys.} \textbf{\bibinfo{volume}{50}},
  \bibinfo{pages}{4831} (\bibinfo{year}{1969}).

\bibitem[{\citenamefont{Allen and Tildesley}(1989)}]{allen_tildesley_book}
\bibinfo{editor}{\bibfnamefont{M.~P.} \bibnamefont{Allen}} \bibnamefont{and}
  \bibinfo{editor}{\bibfnamefont{D.~J.} \bibnamefont{Tildesley}}, eds.,
  \emph{\bibinfo{title}{Computer Simulation of Liquids}}
  (\bibinfo{publisher}{Clarendon Press Oxford}, \bibinfo{address}{Oxford},
  \bibinfo{year}{1989}).

\end{thebibliography}

\end{document}